\documentclass[reprint,pra,superscriptaddress,nofootinbib]{revtex4-2}
\usepackage{amsmath,amssymb}
\usepackage{graphicx}
\usepackage[dvipsnames]{xcolor}
\usepackage[colorlinks=true,linkcolor=blue,citecolor=blue,urlcolor=blue]{hyperref}

\newcommand{\F}{\mathbb{F}_2}
\newcommand{\wt}{\operatorname{wt}}
\begin{document}

\title{Random encoders and expurgation improve 1,873 entries of the Grassl qubit code tables}

\author{Marin Girard}
\affiliation{Virginia Tech, Blacksburg, Virginia 24061, USA}

\date{\today}

\begin{abstract}
I report 192 explicit binary stabilizer codes $[[n,k,d]]$ with $53\le n\le 254$ and $5\le d\le 13$ whose minimum distance exceeds the lower bound listed in Grassl's code tables at the same $(n,k)$: 161 codes by one and 31 by two. Through the standard propagation rules they raise 1,873 entries of the table. For 190 of the codes, a random Clifford encoder is followed by expurgation: every logical operator of weight below the target distance is enumerated, and a few logical operators are promoted to stabilizers so that each low-weight one anticommutes with at least one of them. The other two are obtained by puncturing a code from the table and repairing it in the same way. All 192 distances are exact: the lower bound is certified by exhaustive meet-in-the-middle enumeration of up to $4\times10^{13}$ Pauli operators, and the upper bound by an explicit logical operator of weight $d$. A first-moment count explains the results. Expurgated random codes have about eight more logical qubits than a typical random code of the same length and distance, so they beat the table wherever its best known codes are no better than random. The largest gains, up to nine logical qubits, are at $d\ge9$ and $n\gtrsim170$, where the table often gives a lower bound without an explicit code. The same count predicts improvements at 619 further pairs $(n,d)$. The search was run by AI agents. The codes, their certificates and an independent checker are public.
\end{abstract}

\maketitle

\section{Introduction}

Grassl's tables~\cite{grassl} list, for every length $n\le 256$ and dimension $k$, the best known lower and upper bounds
on the minimum distance of qubit stabilizer codes~\cite{gottesman,crss}. At lengths above 100, the explicit codes
behind the lower bounds are mostly cyclic-type codes, shortened or extended by extra qubits: in a random sample of 26
entries with $n\ge100$ and $5\le d\le13$, 21 come from BCH, cyclic, constacyclic, quasi-cyclic or twisted codes, 2 from
stored generator matrices, and 3 have no explicit code. Over the whole table, more than half of the entries with
$n\ge100$ (15,011 of 27,946 with $k\ge1$) give a lower bound but no explicit code. Random stabilizer codes reach the
quantum Gilbert--Varshamov bound~\cite{fengma}, and over most of the table the best known codes are much better than
that random-code level (Fig.~\ref{fig:codes}).

This note reports that random Clifford codes, improved by expurgation and certified by exhaustive enumeration, beat
the table at 192 coordinates $(n,k)$. The codes have dense generators and are not low-density parity-check codes; their
interest lies in their parameters. Section~\ref{sec:construction} describes the construction,
Sec.~\ref{sec:counting} a counting argument that explains and predicts where it wins, Sec.~\ref{sec:results} the
results, and Sec.~\ref{sec:certification} the distance certification. The work was done in the Genesis project, during
a three-day hackathon at Argonne National Laboratory: AI agents wrote and ran the search, and all heavy computations
ran on the Aurora supercomputer of the Argonne Leadership Computing Facility. Section~\ref{sec:ai} describes the agents'
role.

\section{Construction}\label{sec:construction}

A stabilizer group $S$ with $n-K$ independent generators on $n$ qubits defines an $[[n,K]]$ code. Pauli operators up
to phase are vectors in $\F^{2n}$ with the symplectic form, $N(S)$ denotes the normalizer of $S$, and the logical group
$\mathcal{L}=N(S)/S\cong\F^{2K}$ inherits a nondegenerate symplectic form. The distance is
$d=\min\{\wt(P):P\in N(S)\setminus S\}$. A code is given by its $(n-K)\times 2n$ check matrix $H$ in $[X|Z]$ form,
whose rows are the generators.

\emph{Parent codes.} The parent code is defined by a random Clifford encoder: 16--20 layers of random two-qubit
Clifford gates on random pairs of qubits, applied to $K$ logical input qubits and $n-K$ ancilla qubits in $|0\rangle$.
For the present purpose these parents behave like uniformly random stabilizer codes: over 683 parents, the number of
logical classes of weight below the target distance agreed with the random-code expectation $E$ of Eq.~\eqref{eq:E}
to within about 1\%.

\emph{Expurgation.} Promoting logical operators to stabilizers gives a subcode, the quantum counterpart of
expurgating a classical code~\cite{macwilliams}; it is one of the standard constructions of the
tables~\cite{crss,grassl}. Here the logical operators are chosen to remove every low-weight logical operator. Fix a
target distance $D$ and let $W\subset\mathcal{L}$ be the set of nonzero logical classes that contain a Pauli operator
of weight below $D$. The set $W$ is computed exactly with the enumeration of Sec.~\ref{sec:certification}. Choose an
isotropic subspace $U\subseteq\mathcal{L}$ of dimension $q$ and add representatives of its elements to the stabilizer;
in terms of check matrices, $q$ logical operators are appended to $H$ as new rows. The group $S'=\langle S,U\rangle$
defines an $[[n,K-q]]$ code whose nontrivial logical classes are those in $U^{\perp}\setminus U$, where $U^{\perp}$ is
the symplectic complement of $U$ in $\mathcal{L}$. Hence
\begin{align}
d(S')\ge D \iff W\cap U^{\perp}\subseteq U ,\label{eq:criterion}
\end{align}
that is, every low-weight logical class must anticommute with some element of $U$ or become a stabilizer. In practice
the generators $u_1,\dots,u_q$ of $U$ are chosen so that every $w\in W$ anticommutes with at least one $u_j$; they are
found by a randomized search that uses linear algebra over $\F$ to impose these constraints. With this choice no class of $W$ lies in $U$, so no low-weight logical operator of the parent becomes a stabilizer. This is why the codes are pure (Sec.~\ref{sec:certification}).

\emph{Puncture and repair.} Two of the codes start from a table code instead of a random one. One qubit of
$[[113,73,9]]$ or $[[113,85,7]]$ is punctured, which gives $[[112,74]]$ and $[[112,86]]$ parents with distance at
least 8 and 6. Expurgation with $q=5$ and $q=7$ then yields $[[112,69,9]]$ and $[[112,79,7]]$. This beats the table
only where its $k$ drops by more than $q-1$ when the length decreases by one, which is rare.

\section{Where the method wins}\label{sec:counting}

\emph{The random-code level.} A random Pauli operator commutes with the $n-k$ generators of a random $[[n,k]]$ code
with probability about $2^{-(n-k)}$, so the expected number of nontrivial logical operators of weight below $d$ is
\begin{align}
E(n,k,d)=2^{-(n-k)}\sum_{w=1}^{d-1}\binom{n}{w}3^{w}.\label{eq:E}
\end{align}
A random code typically has distance at least $d$ when $E<1$, that is, when $k<k_{\rm rand}(n,d)$ with
\begin{align}
k_{\rm rand}(n,d)=n-\log_2\sum_{w=1}^{d-1}\binom{n}{w}3^{w}.
\end{align}
This is the random-code level. A rigorous existence bound of this form, the quantum Gilbert--Varshamov bound at finite length, was proved for pure stabilizer codes by Feng and Ma~\cite{fengma}. Each extra logical qubit doubles $E$, so it is natural to measure codes on a logarithmic scale. The excess
\begin{align}
X(n,k,d)=\log_2 E(n,k,d)=k-k_{\rm rand}(n,d)
\end{align}
counts how many logical qubits a code has above the random-code level. Two codes with the same $n$ and $d$ differ in
$k$ by exactly the difference of their excesses.

\emph{What expurgation reaches.} Expurgation can beat a random code because $U$ is chosen rather than drawn. How far
it can go follows from a first-moment count, in four steps.

\emph{Step 1: what must be removed.} Suppose the goal is a final code $[[n,k,D]]$, reached from a parent with
$K=k+q$ logical qubits by adding $q$ stabilizers. The parent has $q$ more logical qubits than the final code, so its
excess is larger by $q$, and its number of low-weight logical classes is, by Eq.~\eqref{eq:E},
\begin{align}
m=|W|\approx 2^{\,X(n,K,D)}=2^{\,q}\,2^{\,X(n,k,D)}.
\end{align}
All $m$ classes must be removed by the $q$ new stabilizers.

\emph{Step 2: a random choice of $U$ is no better than a random code.} A class $w\notin U$ survives if it lies in
$U^{\perp}$, that is, if it commutes with every generator $u_1,\dots,u_q$ of $U$. Whether $w$ commutes with $u_j$ is
one bit, and for a randomly chosen $U$ each of these $q$ bits is zero with probability about $1/2$, so $w$ survives with
probability about $2^{-q}$. The expected number of survivors is then $m\,2^{-q}=2^{X(n,k,D)}$: exactly the number of
low-weight logical operators of a random $[[n,k]]$ code. This is no coincidence, since a randomly chosen $U$ simply
produces another random code. Treating the survivals of the $m$ classes as independent, the number of survivors is
approximately Poisson distributed, and a random $U$ removes every class with probability
\begin{align}
P_{\rm good}=\bigl(1-2^{-q}\bigr)^{m}\approx e^{-m2^{-q}}=e^{-2^{X(n,k,D)}}.
\end{align}
This is tiny once $X$ exceeds a few units: for $X=8$ it is $e^{-256}\approx2^{-369}$.

\emph{Step 3: there are many choices of $U$.} An isotropic subspace of dimension $q$ is spanned by $q$ mutually
commuting logical operators. The first can be any of the $2^{2K}-1$ nonzero classes. The $j$-th must commute with the
previous $j-1$, and each of these conditions halves the number of candidates; it must also lie outside their span. This
leaves about $2^{2K-j+1}$ choices, so there are about $\prod_{j=1}^{q}2^{2K-j+1}=2^{2Kq-q(q-1)/2}$ ordered choices.
Each subspace is counted once for each of its bases, and a $q$-dimensional space has about $2^{q^2}$ bases, so the number
of subspaces is about
\begin{align}
2^{\,2Kq-q(q-1)/2-q^{2}}\approx2^{\,2Kq-\frac{3}{2}q^{2}},
\end{align}
up to a factor $2^{O(q)}$. Choosing $U$ therefore offers about $2Kq-\frac{3}{2}q^{2}$ bits of freedom.

\emph{Step 4: weigh the freedom against the cost.} Multiplying the number of subspaces by the probability that one of
them works gives the expected number of subspaces that satisfy Eq.~\eqref{eq:criterion},
\begin{align}
\mathcal{N}\approx 2^{\,2Kq-\frac{3}{2}q^{2}}\,\bigl(1-2^{-q}\bigr)^{m}.\label{eq:N}
\end{align}
If $\mathcal{N}\gg1$, good subspaces should exist; if $\mathcal{N}\ll1$, they almost certainly do not. In base-2 logarithms, using
$\log_2 e^{-x}=-x/\ln2$ and Step 2,
\begin{align}
\log_2\mathcal{N}\approx\Bigl(2Kq-\tfrac{3}{2}q^{2}\Bigr)-\frac{2^{X(n,k,D)}}{\ln2}.
\end{align}
The first term is the freedom in choosing $U$; the second is the cost, in bits, of removing every low-weight class.
Requiring $\log_2\mathcal{N}\gtrsim0$ gives $2^{X(n,k,D)}\lesssim\ln2\,(2Kq-\frac{3}{2}q^{2})$, and taking the
logarithm once more,
\begin{align}
X(n,k,D)\lesssim \log_2\!\Bigl[\ln 2\,\Bigl(2Kq-\tfrac{3}{2}q^{2}\Bigr)\Bigr].\label{eq:bound}
\end{align}

\emph{Reading the bound.} The cost of removing the low-weight classes grows exponentially with the excess, as
$2^{X}$, while the freedom grows only like $2Kq$. The excess that expurgation can reach is therefore only logarithmic in
$K$ and $q$. For example, $[[180,131,9]]$ came from a $K=138$ parent with $q=7$: the freedom is
$2\cdot138\cdot7-\frac{3}{2}\cdot49\approx1860$ bits, and Eq.~\eqref{eq:bound} gives $X\lesssim\log_2(0.69\times1860)
\approx10.3$, while the code reached $X=8.1$. Increasing $q$ raises the freedom only linearly, while the number of
classes that must be enumerated and removed, $m=2^{X+q}$, doubles with each extra stabilizer; in practice $q\le8$.

Over the parameters used here the right-hand side of Eq.~\eqref{eq:bound} lies between 5.8 and 11.1 (median 10.2).
For the 160 codes whose $K$ and $q$ were recorded, the achieved $X$ is 0.8 to 3.0 below it (median 2.3 below). A
shortfall is expected: a first-moment estimate is optimistic, and the search does not examine every subspace. Because
$K$ and $q$ enter only through a logarithm, the new codes all sit at nearly the same excess: 90\% of them have
$6.1\le X\le 8.9$, with median 7.7.

\emph{A calibrated predictor.} For a practical prediction we take, for each parent dimension $K$, the smallest $q$
with $\log_2\mathcal{N}\ge0$ in Eq.~\eqref{eq:N} (using the exact $\log_2(1-2^{-q})$), add one to it, keep $q\le8$,
maximize $K-q$ over $K$, and subtract one. These two offsets were fixed on early runs. At the 192 new codes the
predictor gives the achieved $k$ exactly for 147 of them and within one for 178. Before the last searches were run it
predicted $k=106$, 121 and 54 for $(n,d)=(172,12)$, $(188,12)$ and $(120,13)$; the searches produced exactly these
values, where the table has 98, 113 and 48.

\begin{figure*}[t]
\centering
\includegraphics[width=\textwidth]{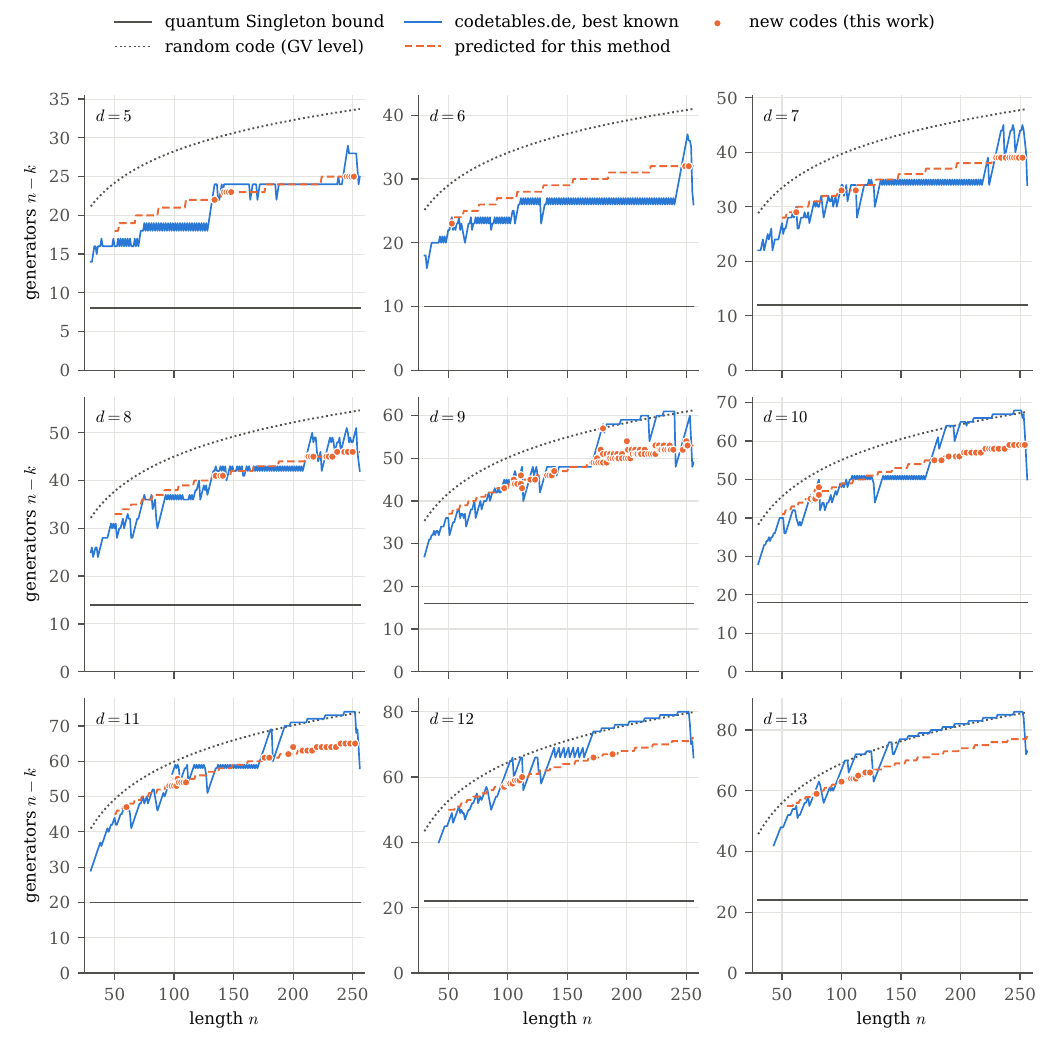}
\caption{Number of stabilizer generators $n-k$ needed at length $n$ for distance $d$; lower is better, and at fixed
$(n,d)$ a vertical step of one is one logical qubit. Solid gray: quantum Singleton bound $n-k\ge2(d-1)$. Dotted gray:
random-code (Gilbert--Varshamov) level $n-k_{\rm rand}(n,d)$. Blue: best code in Grassl's table~\cite{grassl}, read on
4 November 2026. Dashed orange: the calibrated predictor of Sec.~\ref{sec:counting} for random encoders with
expurgation. Orange dots: the 192 new codes. Where the blue line lies above the orange dashed one, the predictor expects random
encoders with expurgation to improve the table.}
\label{fig:codes}
\end{figure*}

\emph{Where the table has no explicit codes.} Figure~\ref{fig:codes} compares, for each distance, the number of
generators $n-k$ of the table's best codes with the random-code level, the predictor and the new codes. Where
cyclic-type codes exist, the table's codes are far better than random and expurgated random codes cannot compete. Over
long stretches of $n$, however, the table is only as good as a random code, and these stretches are exactly where it
lists a lower bound but no explicit code. For 70 of the new codes, the table's best entry at the same $n$ and $d$ has no
construction, and each of these 70 entries lies within one logical qubit below the random-code level ($-1<X<0$), as an
existence bound of Gilbert--Varshamov type would. The new codes gain 8.2 logical qubits there on average, against 3.5
where the table has an explicit code. For example, the $d=11$ codes every four qubits from $n=208$ to 252 beat the table
by 5 to 9 logical qubits at every length from 208 to 254, leaving the predictor at most three more there; at $n=255$ a
BCH code takes over.

\emph{Predictions.} Applying the predictor to every $n\le256$ and $5\le d\le13$ gives 619 pairs $(n,d)$ where it
expects a code better than both the table and the present codes. For 378 of them the expected gain is one to three
logical qubits, spread over all distances. For 192 it is six to nine, almost all at $d=12$ ($n\approx178$--252) and
$d=13$ ($n\approx138$--253), in stretches where 178 of these 192 table entries have no explicit code. Certifying one
such code requires enumerating $5\times10^{11}$ to $2.5\times10^{14}$ Pauli operators, up to six times the largest run
reported here. The predicted line is drawn in Fig.~\ref{fig:codes}, and the full list is in the repository.

\section{Results}\label{sec:results}

Table~\ref{tab:summary} summarizes the codes by distance. At its own $(n,k)$, each code exceeds the table's lower bound,
161 by one and 31 by two. At fixed $(n,d)$ the gain in $k$ ranges from 1 to 9, and 66 codes gain at least 8. Examples
are $[[180,131,9]]$, $[[212,149,11]]$, $[[172,106,12]]$, $[[188,121,12]]$, $[[124,58,13]]$ and $[[112,48,13]]$, where
the table has $k=122$, 140, 98, 113, 51 and 40 at the same $n$ and $d$.

The propagation rules $[[n,k,d]]\to[[n+1,k,d]]$ and $[[n,k,d]]\to[[n,k-1,d]]$~\cite{crss} carry each code to longer
lengths and smaller dimensions. Together they raise 1,873 table entries with $k\ge1$, 1,748 by one and 125 by two; 485
of these entries had no explicit code. The comparison uses the table as read on 4 November 2026.

\begin{table}[t]
\caption{The new codes by distance $d$: number of codes, range of $n$, how many exceed the table's lower bound at their
own $(n,k)$ by one or by two, and the largest gain in $k$ at fixed $(n,d)$ (table value in parentheses).}
\label{tab:summary}
\begin{ruledtabular}
\begin{tabular}{rrcrrl}
$d$ & codes & $n$ & $+1$ & $+2$ & largest gain in $k$ \\
\hline
5  &   9 & 134--251 &  9 &  0 & $+3$: $[[245,220,5]]$ (217) \\
6  &   3 &  53--252 &  3 &  0 & $+4$: $[[252,220,6]]$ (216) \\
7  &  13 &  62--252 & 13 &  0 & $+6$: $[[252,213,7]]$ (207) \\
8  &  14 & 135--250 & 14 &  0 & $+4$: $[[215,170,8]]$ (166) \\
9  &  73 &  97--251 & 73 &  0 & $+9$: $[[180,131,9]]$ (122) \\
10 &  28 &  74--254 & 13 & 15 & $+9$: $[[213,156,10]]$ (147) \\
11 &  31 &  60--252 & 18 & 13 & $+9$: $[[212,149,11]]$ (140) \\
12 &  12 &  97--188 & 11 &  1 & $+8$: $[[104,46,12]]$ (38) \\
13 &   9 &  78--124 &  7 &  2 & $+8$: $[[112,48,13]]$ (40) \\
\hline
all & 192 & 53--254 & 161 & 31 & \\
\end{tabular}
\end{ruledtabular}
\end{table}

\section{Distance certification}\label{sec:certification}

\emph{Upper bound.} For every code the repository lists a logical operator of weight exactly $d$, found by a
random-window search. Checking it takes a few lines of linear algebra: it commutes with every generator and is not in
the stabilizer group.

\emph{Lower bound.} Let $b=\lfloor (D-1)/2\rfloor$ and $a=D-1-b$. Every Pauli operator $P$ of weight at most $D-1$ splits as $P=AB$ with disjoint supports, $\wt(B)\le b$ and $\wt(A)\le a$. If $P\in N(S)$ then $A$ and $B$ have the same syndrome. It therefore suffices to enumerate all Pauli operators of weight at most $b$, together with all of weight exactly $a$ when $a=b+1$, find every pair with equal syndromes, and check each product exactly: it is either a stabilizer, which is harmless, or a nontrivial logical operator, which refutes $d\ge D$. The number of syndrome evaluations is
\begin{align}
\sum_{w=0}^{b}\binom{n}{w}3^{w}+[a>b]\binom{n}{a}3^{a},\label{eq:cost}
\end{align} up to $4.2\times10^{13}$ for $[[188,121,12]]$. Every run records how many Pauli operators it visited, and in every complete run that count equals this closed form.

\emph{Purity.} All 192 codes are pure: the stabilizer group contains no element of weight below $d$ other than the identity. The minimum weight of the whole normalizer $N(S)$, stabilizers included, is therefore $d$. For the random codes this follows from the construction. No low-weight logical operator of the parent is promoted to a stabilizer (Sec.~\ref{sec:construction}), and the stabilizer group of a random parent contains on average about $m\,4^{-K}$ elements of weight below $D$, which is negligible. Purity was also checked directly for every code.

\emph{Implementations.} The two certifiers written during the search certified 190 of the codes on Aurora. A second,
independent certifier, written afterwards with an AI assistant but sharing no code with the search, was tested against
brute-force distances and certified 32 codes on a laptop, including the remaining two. Every code thus has at least one
complete certificate, and 30 have certificates from two independent implementations. Re-checking the codes with
$d\ge10$ with the independent certifier needs a cluster.

\section{Outlook}

Exhaustive certification is what limits this approach. Its running time grows like $n^{d/2}$, and so does its memory:
the half of the enumeration that must be stored has $\sum_{w\le b}\binom{n}{w}3^w$ entries, $3.3\times10^{12}$ for
$[[124,58,13]]$, or about 26~TB at 8 bytes per entry, which must be spread over many nodes or processed in passes. At
$n\approx250$ and $d\approx13$ it already needs a supercomputer. Codes built as tensor networks with small cuts avoid
this. Their weight enumerators, and hence their distances, can be computed by contracting the network, at a cost
exponential in the cut size rather than in $n$~\cite{enumerators}. Random encoders have no small cuts. A family of
encoders that is random within blocks but joined through small cuts might keep most of the excess of random codes while
allowing efficient distance computation, and could reach greater lengths. Quantum-LEGO constructions~\cite{lego} did
not contribute to the findings here but are a natural framework for such a search.

\section{Role of AI agents}\label{sec:ai}

AI agents did most of the work under my direction: they retrieved and parsed the code tables and the literature
(including the table codes used as starting points for the two punctured codes), wrote the search programs and the
original certifiers, managed the runs on Aurora (splitting searches and certifications into batches, submitting the
jobs, and collecting and reassembling the results), and produced the figures. I chose which constructions to pursue,
allocated the compute, and decided what to verify and how. This note was drafted with an AI assistant and then edited
and checked by me. Every code is given explicitly, its distance upper bound is a short linear-algebra check, and its
lower bound has a complete certificate whose size matches the closed form.

\emph{Other approaches.} The following alternatives were also tried; none did better than random parents. These are
results about the implementations and parameters tried, not general no-go statements.
\begin{itemize}\setlength{\itemsep}{0pt}\setlength{\parskip}{0pt}
\item \emph{Structured parents.} Quasi-cyclic and translation-invariant encoder circuits instead of random ones. At
the parameters tested they had as many low-weight logical classes as random parents.
\item \emph{Codes assembled from small pieces.} Tensor networks of random few-qubit codes, and concatenation trees.
They were useful as controls but did not reach the table.
\item \emph{Block expansion.} Parents built from $[[4,2,2]]$ code blocks. At $n=32$ they had about five times more
low-weight logical classes than random parents, and none of four led to a certified code, against four of four for
random parents.
\item \emph{Editing published codes.} Removing generators from a table code and adding others. Removing generators
exposed logical operators of weight one or two.
\end{itemize}

\section*{Data availability}
The public repository~\cite{repo} contains, for every code, the stabilizer matrix in $[X|Z]$ form, a logical operator of
weight $d$, and the records of its lower-bound runs: which certifier ran, how many Pauli operators it enumerated
(compared with Eq.~\eqref{eq:cost}) and the result. It also contains the comparison with the table, the predictions of
Sec.~\ref{sec:counting}, and two checking programs: a short pure-Python script that checks each code and its weight-$d$
logical operator, and the independent lower-bound certifier with its tests. The project initially aimed to design codes
as quantum-LEGO tensor networks~\cite{lego}, and the random encoders were generated in that framework, so the repository
also gives, for each code, the tensor-network recipe from which it can be rebuilt. The stabilizer matrices alone are
enough to use and check the codes.

\begin{acknowledgments}
I thank ChunJun Cao for supervision and discussions, and Argonne National Laboratory for hosting the Genesis hackathon. I thank Markus Grassl for his comments on the first version of this note. This work utilized resources made available by the U.S. Department of Energy's Genesis Mission. This work was supported by the U.S. Department of Energy, Office of Science, Office of Advanced Scientific Computing Research, as part of the Genesis Mission National Science and Technology Challenges, under Award Number DE-SC0026692.
\end{acknowledgments}

\end{document}